# Atomistic manipulation of reversible oxidation and reduction in Ag by electron beam

*Huaping Sheng,[1,2] He Zheng,[2] Lifen Wang,[1] Shuangfeng Jia,[2] Huihui Liu,[2] Maria K.Y. Chan,[1] Tijana Rajh,[1] and Jianbo Wang[2,3]\* Jianguo Wen,[1]\**

[1]Center for Nanoscale Materials, Nanoscience and Technology, Argonne National Laboratory, Argonne, IL 60439, USA

[2]School of Physics and Technology, Center for Electron Microscopy, MOE Key Laboratory of Artificial Micro- and Nano-structures, and Institute for Advanced Studies, Wuhan University, Wuhan 430072, China

[3]Science and Technology on High Strength Structural Materials Laboratory, Central South University, Changsha 410083, China

**ABSTRACT**: Employing electrons for direct control of nanoscale reaction is highly desirable since it provides fabrication of nanostructures with different properties at atomic resolution and with flexibility of dimension and location. Here, applying *in situ* transmission electron microscopy, we show the reversible oxidation and reduction kinetics in Ag, well controlled by changing the dose rate of electron beam. Aberration-corrected high-resolution transmission electron microscopy observation reveals that O atoms are preferably inserted and extracted along the {111} close-packed planes of Ag, leading to the nucleation and decomposition of nanoscale $Ag_2O$ islands on the Ag substrate. By controlling electron beam size and dose rate, we demonstrated fabrication of an array of 3 nm $Ag_2O$ nanodots in an Ag matrix. Our results open up a new pathway to manipulate atomistic reaction with electron beam towards the precise fabrication of nanostructures for device applications.

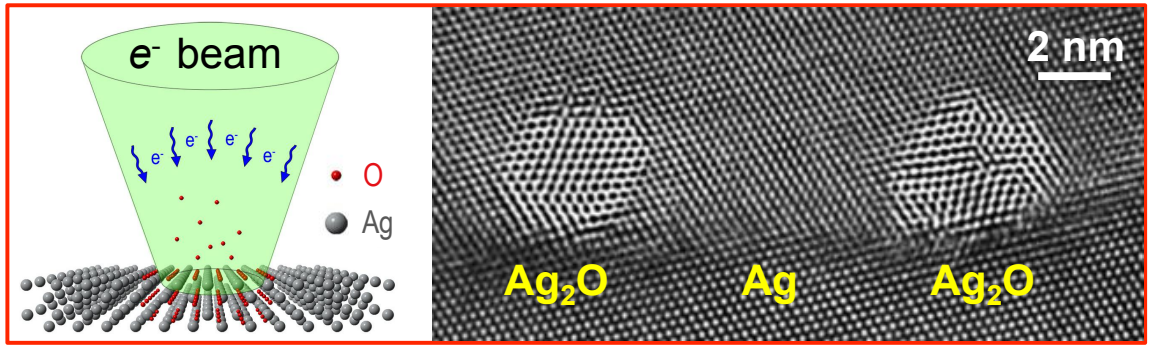

The fabrication of ordered nanoscale structures with desired property, dimension, and location has attracted considerable scientific and commercial attention because of their potential utilization in electronic, optical, micromechanical, and quantum devices and sensors.[1–3] There are two major approaches to fabricate nanometer-sized structures: "top-down" approach using nanolithography, thin film deposition and etching techniques, and "bottom-up" approach using chemical synthesis or self-assembly.[4,5] Employing electron beam for nanofabrication is highly desirable since it provides both the highest resolution and location flexibility. So far, electron beam nanofabrication in a microscope mainly consists of direct electron beam writing such as electron-beam induced deposition (EBID)[6] and electron beam sculpting.[7,8] EBID is a promising technique to combine "top-down" and "bottom-up" approaches using a focused electron beam to decompose organic or metalorganic gases but it reaches a 15 nm resolution limit even though a much smaller electron probe is used.[9] Electron beam sculpting is a technique to remove atoms or change material crystallinity into amorphous with electron beam irradiation. Recently, Jesse et al demonstrated atomic-level sculpting of crystalline oxides using a focused electron beam with a scanning transmission electron microscope (STEM) taking the advantage of the significant advance in aberration correction technology enabling formation of a sub-angstrom probe with a probe corrector.[10–12] In addition to these two major approaches, it is highly desirable to develop a new approach to direct control of nanoscale reaction using electron beam due to its potential utility in the fabrication of nanoscale devices.

Both silver (Ag) and silver oxide ($Ag_2O$) are promising materials showing broad potential applications in photo-activated fluorescence,[13] strong plasmon-resonant optical scattering,[14] surface passivation,[15] low-temperature catalyzing[16], antibacterial ability,[17] etc.

Being a p-type semiconductor with a bandgap of 1.2 ~ 3.1 eV[18–21], a Ag-Ag$_2$O metal-semiconductor contact should be possibly fabricated by locally oxidizing Ag. Since most of these properties involve both metallic Ag and Ag$_2$O simultaneously, a deep understanding and further controlling on the redox of Ag-Ag$_2$O conversion are of significant scientific and technological importance. Plentiful works have been done to investigate the interactions between Ag and O.[22–35] A conspicuous finding is that Ag is subjected to oxidation very easily when exposed to atomic O, while remains stable in molecular O environments.[27,30,31,36] A recent work by Sun et al. shows that the oxidation of Ag can be induced by ionizing the O$_2$ using electron-beam irradiation in an environmental gas cell, and new Ag will nucleate under intense electron-beam irradiation above a limit.[37] Even though similar studies have been performed regarding the redox dynamics of metals and impressive results have been obtained by Zhou et al,[38–41] details corresponding to Ag remain elusive. Also, the precise control of the equilibrium is yet to achieve.

In this work, we report the controllable oxidation-reduction of Ag-Ag$_2$O at the atomic level by manipulating the electron beam inside a transmission electron microscope. The atomistic reaction dynamics were directly captured by high resolution transmission electron microscopy (HRTEM), enabling us to understand and monitor the reaction dynamics at atomic level. Due to the lower contraction of oxygen in a microscope, the redox reaction is slowed dramatically, enabling us to control the reaction dynamics easily. By utilizing the electron-beam, we demonstrated the nano-patterning of Ag$_2$O nanowire and a 3 nm nano-arrays in an Ag matrix.

Thin Ag foils (99.95%, purchased from the Alfa Aesar) were punched into 3 mm disks and ion milled to electron transparency using 4 kV Ar$^+$ ions at low angles of 5 degrees.

As prepared Ag TEM specimen was directly loaded into a FEI Titan TEM equipped with image corrector to correct both spherical and chromatic aberrations, working at 200 kV.

Firstly, it is found that pure Ag can be easily oxidized under the electron-beam irradiation with a low electron beam intensity as signified by the nucleation of $Ag_2O$ island (Figure 1a-1b), indicating that even in the high vacuum environment (~$10^{-7}$ Torr pressure), the remained ionized oxygen can still induce the oxidation of pure Ag. Interestingly, with the increase of electron-beam intensity by focusing the electron-beam, the formed $Ag_2O$ islands could be fully reduced back to Ag (Figure 1c), signifying a reversible oxidation-reduction dynamics dependent on the electron-beam intensity, i.e. the dose rate of electron-beam. See also the experimental data displayed as Figure 1d-f. The dose-rate dependent growth (oxidation) and decomposition (reduction) rate of $Ag_2O$ islands were shown in Figure 1g. Specifically, the $Ag_2O$ island kept growing under electron-beam irradiation with a dose rate of ~$10^5$ e $Å^{-2}$ $s^{-1}$; when the dose rate was increased to ~$10^6$ e $Å^{-2}$ $s^{-1}$, the $Ag_2O$ island started to shrink and reverted back to Ag (figure 1g). It is believed that both the oxidation and reduction take place simultaneously during the whole process, and the apparent dose rate-dependent growth or decomposition of $Ag_2O$ should originate from the competition between these two reactions. There exists an equilibrium electron-beam dose rate condition (~ $5 \times 10^5$ e $Å^{-2}$ $s^{-1}$) the oxidation/reduction rates are equivalent. A growth or decomposition of $Ag_2O$ island depends on electron-beam dose rate is lower or higher than the equilibrium electron-beam dose rate.

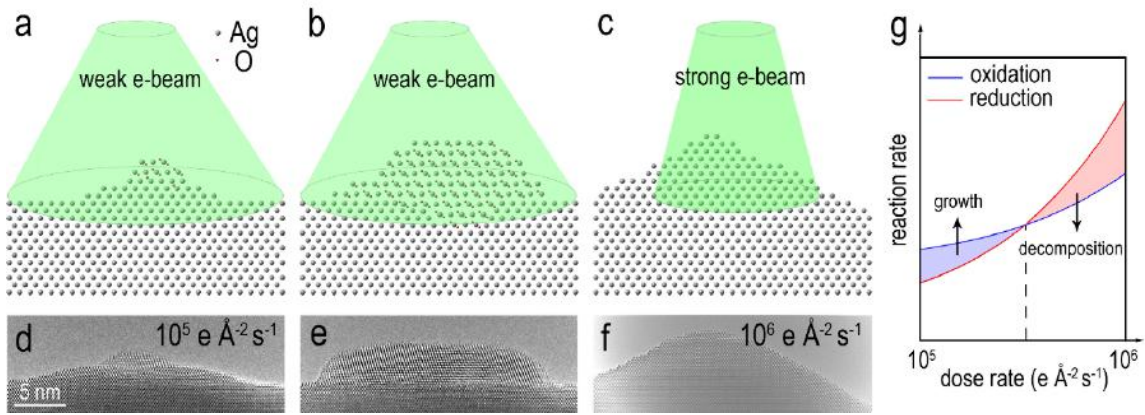

**Figure 1.** Schematic drawing of the reversible oxidation (a-b) and reduction (c) in Ag-Ag$_2$O. (d-f) HREM images showing the corresponding processes. (g) The electron-beam dose rate dependent growth and decomposition of Ag$_2$O island.

When the newly formed Ag$_2$O grain is subjected to further electron-beam irradiation, it would either grow bigger by depleting the Ag substrate, or decompose, leading to the expansion of the Ag substrate as a result. The growth and decomposition of as formed Ag$_2$O grain are fully reversible depending on the dose rate of the electron-beam, with the premise that Ag$_2$O doesn't decompose completely and a small nucleus remains on the substrate (see the growth-decomposition cycles in supporting information video file 1). As an example, several time-elapsed images are extracted from the *in situ* recorded video and displayed in figure 2. Setting figure 2a as the starting point (0 s), a small Ag$_2$O is attached to the substrate. Under the electron-beam irradiation of ~$10^5$ e Å$^{-2}$ s$^{-1}$, the Ag$_2$O grain kept growing until the electron-beam dose rate was increased to ~$10^6$ e Å$^{-2}$ s$^{-1}$ at 55 s (figure 2a-2c), after which the grain size started to decrease (figure 2d-2f). At the time point of 104 s, both the size and shape of the Ag$_2$O grain resembled those at 0 s, proving that the oxidation behavior can be reversed and consequently an equilibrium between Ag and Ag$_2$O is achievable by controlling the electron-beam without introducing any apparent changes or defects. The number of electrons needed to form one oxygen atom in $Ag_2O$ is estimated from Fig. 2a-2c to be around ~$3 \times 10^6$.

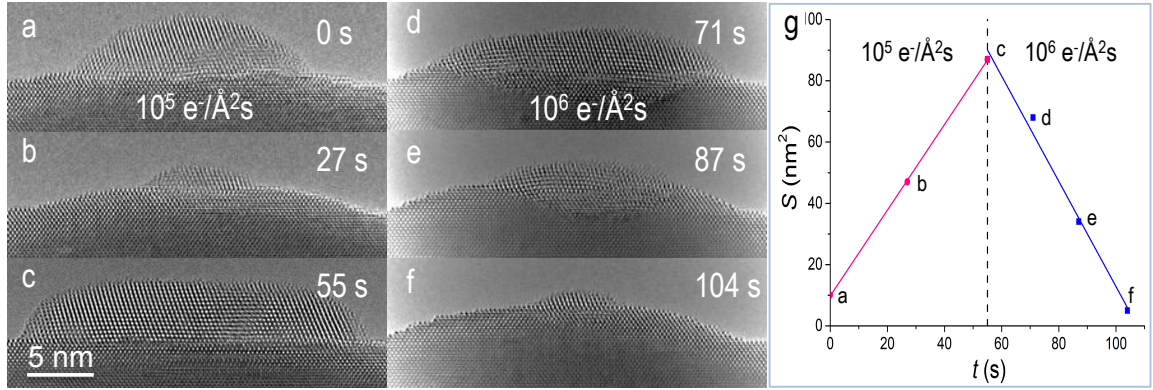

**Figure 2.** Time elapsed images showing a growth-decomposition cycle of the $Ag_2O$ grain attached to the Ag substrate. (a-c) The $Ag_2O$ keeps growing under electron-beam irradiation with a dose rate of ~$10^5$ e $Å^{-2}$ $s^{-1}$. (d-e) Decomposition of the $Ag_2O$ crystal under electron-beam irradiation with a dose rate of ~$10^6$ e $Å^{-2}$ $s^{-1}$. (f) Linear fitting of the $Ag_2O$ grain's projected area ($S$) versus the irradiation time ($t$).

The projected area $S$ ($nm^2$) of the $Ag_2O$ grain at each time point $t$ (s) was estimated and plotted in figure 3f (See also the estimated data in Table S1 in supporting information). Interestingly, linear correlations were found between the $Ag_2O$ grain's area and irradiation time, both in the growth process and decomposition process, similar to high temperature oxidation work by Zheludkevich *et al.*[27] and work by Li *et al.*[42] For figure 2a-2c, the curve is fitted to be $S = 9.73t + 1.40$, indicating an apparent growth speed of 9.73 $nm^2$ $s^{-1}$ of the $Ag_2O$ grain when electron-beam dose rate is about $10^5$ e $Å^{-2}$ $s^{-1}$, while for figure 2d-2f, the fitting result is $S = -1.72t + 184.69$, indicating an apparent decomposition speed of 1.72 $nm^2$ $s^{-1}$ of the $Ag_2O$ grain when electron-beam dose rate is about $10^6$ e $Å^{-2}$ $s^{-1}$. The linear growth and decomposition of the $Ag_2O$ grain should provide some clues toward the underlying mechanism of the oxidation-reduction dynamics stimulated by electron-beam irradiation. A brief explanation for the electron-beam dose rate dependence is the competition between the electron-beam ionization induced oxidation of Ag and electron stimulated desorption (ESD) induced reduction of $Ag_2O$.

The detailed atomistic oxidation and reduction reactions are presented in Figure 3. The formation of a typical $Ag_2O$ island indicates the outward diffusion of Ag atoms which can react with the oxygen ions at the island surface (adatom process). As displayed in the HRTEM image in figure 3a, a small island (colored in red) nucleated on the Ag substrate (colored in blue), as distinguished by the different orientations and $d$-spacings (2.8 Å compared to 2.4 Å) of their lattice planes. The corresponding fast Fourier transformation (FFT) images from squared regions 1 (figure 3b) and 2 (figure 3c) can be indexed based on the $Ag_2O$ and Ag crystal structures, respectively. Even though different oxide species ($Ag_2O$, $AgO$, $AgO_2$, $AgO_3$, etc.) have been reported in the studies of Ag oxidation,[30,37] $Ag_2O$ is the only oxidation product observed in our work, consistent with the previous experimental results of Ag oxidation performed in a low oxygen partial pressure environment similar to a TEM chamber.[27] Moreover, the dominant orientation relationship (OR) between the $Ag_2O$ island and Ag substrate is $<110>_{Ag}//<110>_{Ag_2O}$ and $\{111\}_{Ag}//\{002\}_{Ag_2O}$ based on tens of reaction events. Meanwhile, besides the adatom process mentioned above, the oxidation of Ag at the Ag-$Ag_2O$ interface is also observed, signifying the inward diffusion of oxygen atoms (See Figure S1 in supporting information). Interestingly, twice (Figure S1c) and three times (Figure S1d) structural modulations along the $[1\bar{1}1]$ direction are found in the oxidized Ag, indicating the periodic intercalation of oxygen atoms (Figure 3d). Further work is required to reveal the detailed atomic-scale reaction mechanisms.[43–45]

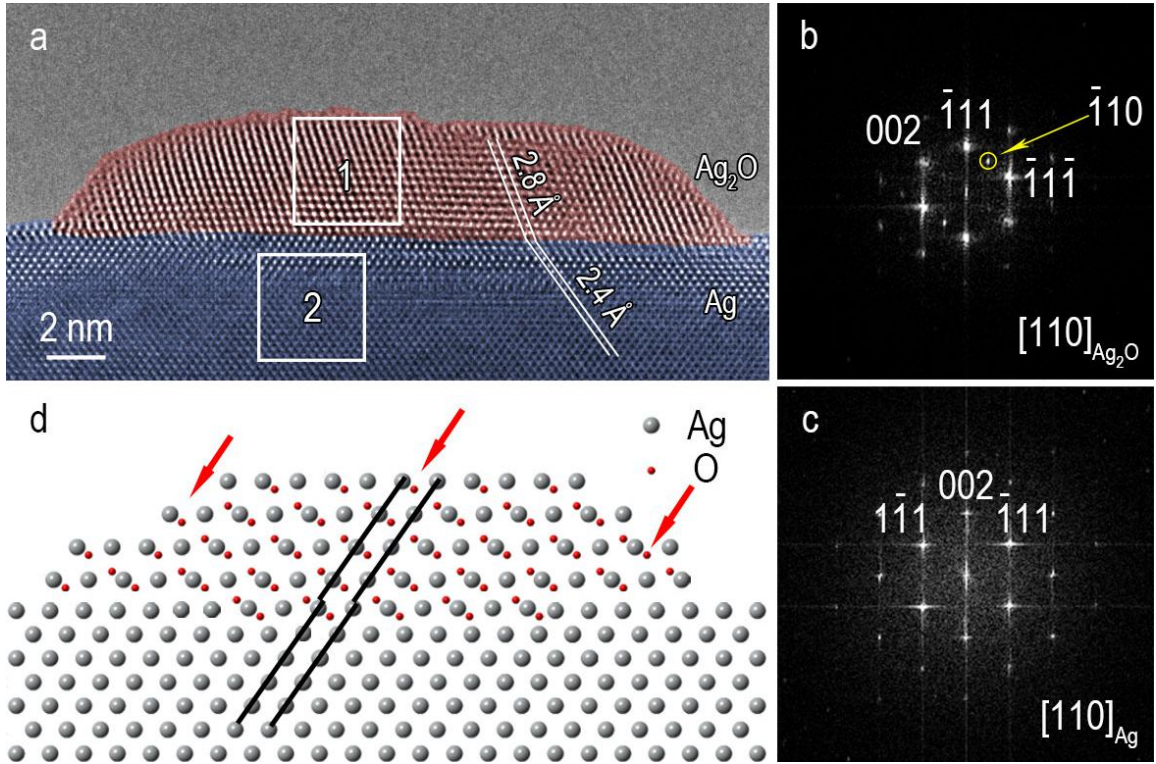

**Figure 3.** Atomic-scale details of the Ag-Ag$_2$O interface. (a) HREM image showing a Ag$_2$O grain (pink) attached to the Ag substrate (blue). (b, c) FFT images of the top Ag$_2$O grain (square 1) and the bottom Ag substrate (square 2), respectively. The marked spot indicates the typical ($\bar{1}$10) spot of Ag$_2$O. (d) Atomic model showing the configuration of Ag/Ag$_2$O interface.

As stated above, the oxidation and reduction rates are controllable by simply controlling the irradiation dose rate, which may prove essential in advancing a range of nanotechnologies in quantum devices fabrication, such as the introduction of metal (Ag) – semiconductor (Ag$_2$O) Schottky junction diodes. The control of oxidation/reduction kinetics with atomic-scale precision is further demonstrated in Figure 4, whereas the electron-beam was made into a spindle shape and maintained at a dose rate of ~$10^5$ e Å$^{-2}$ s$^{-1}$ (Figure 4a). Evidently, after 40 minutes' irradiation, a new phase with exactly the same shape and size as the electron-beam was observed within the irradiated area (figure 4b), which means that the reaction was precisely confined. From the X-ray energy dispersive spectroscopic (EDS) elemental mapping, it is obvious that the irradiated area has accumulated additional Ag and O compared to the un-irradiated region (figure 4c, 4d), illustrating again the oxidation of Ag. Also, the higher content of Ag indicates the migration of Ag atoms from adjacent areas, consistent with the observation shown above ($Ag_2O$ island extruded from the Ag substrate). The steep decreases at the vicinity of metal-metal oxide boundaries in the line-scan data (illustrated by the dashed lines in figure 4e) demonstrate that the oxidation was well confined within the irradiated area (~35 nm wide).

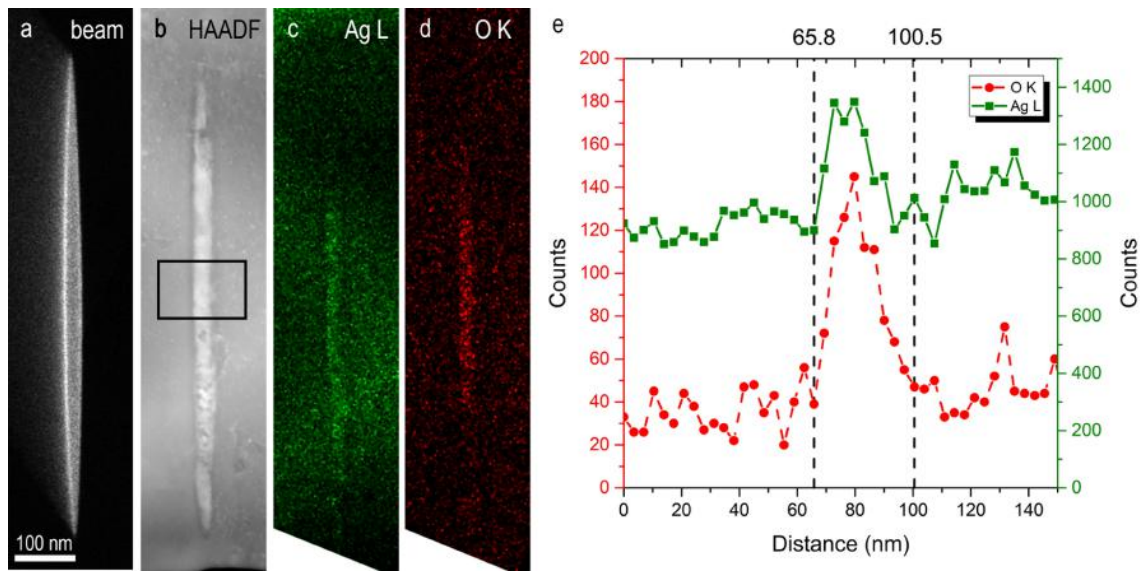

**Figure 4.** (a) BF image showing the shape of electron-beam. (b) HAADF image of the Ag substrate after 40 minutes' irradiation. (c, d) Corresponding EDS mapping of Ag and O elements (e) Linescan profile of the boxed area in (b) extracted from EDS mapping.

Taking another step further, patterned oxidation has also been tested to verify the reliability of the electron-beam irradiation as a nanofabrication technique. electron-beam was focused into a nanometer probe on the sample and programmed to hold up at each preconfigured spot for a fixed time. Figure 5a shows a 3×1 nano-array fabricated in our preliminary experiments. Figure 5b is the enlarged picture of spot 2 in figure 5a and figure 5c is the corresponding FFT image. Again, the newly formed structure was determined to be $Ag_2O$ and possessed the same OR with respect to the Ag substrate, as stated above, demonstrating that the electron-beam irradiation controlled oxidation can be effectively confined within an area down to 3 nm. However, due to the confinement from the Ag substrate, distortions were observed in the newly formed $Ag_2O$ grains because of its larger lattice parameters. As we can see in figure 5a, Ag has been well oxidized into $Ag_2O$ under electron-beam irradiation at spot 1 and 2, while for spot 3, the irradiation process went too far thus leaving a hole on the Ag substrate. The different results for these 3 spots, after exactly the same electron-beam irradiation, originated from the varied thickness of the Ag substrate, which could possibly be eliminated by a dynamically monitor the reaction status and stop the irradiation once the substrate has been well oxidized.

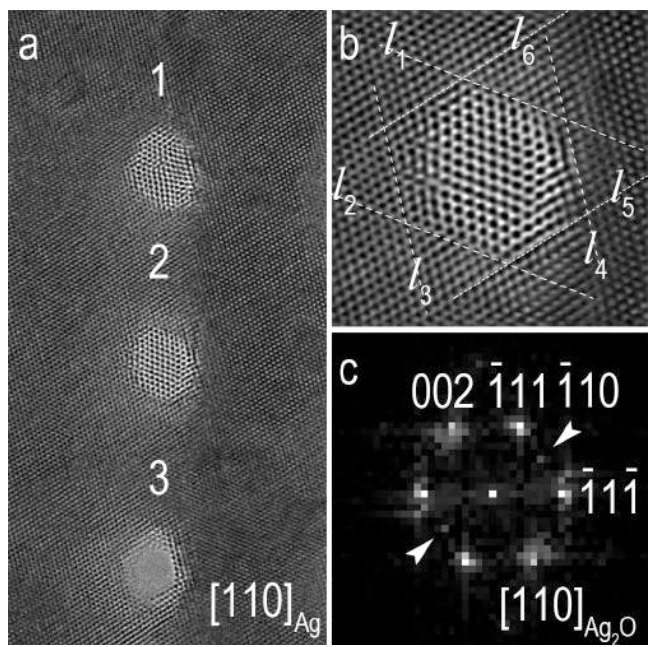

**Figure 5.** Preliminary results on the patterned oxidation of Ag. (a) HREM image showing a Ag$_2$O grain array fabricated by controlling electron-beam irradiation. (b) Zoomed in image of spot 2 in (a) showing the detailed configuration of Ag$_2$O grain on Ag substrate. (c) FFT image of (b) demonstrating the existence of Ag$_2$O crystal and its orientation.

It is worth noting that the specimen was held in the microscope for more than 12 hours before introducing any electron-beam irradiation, and no oxides were observed on the specimen. In contrast, after electron-beam irradiation with a weak beam intensity, the Ag$_2$O structures were observed within the irradiated area (See figure S2 in supporting information), in agreement with the common knowledge that noble metal Ag is resistant to oxidation by molecular oxygen but will be easily oxidized by atomic oxygen.[27,30,31,36,37]. When electron-beam is turned on, the residual O$_2$ molecules in the chamber and those adsorbed on the Ag surface[23,30,42,46] could be excited into active species including O, O$^-$, O$_2^+$, O$^{2-}$, etc,[37] which react with the Ag substrate immediately. When a primary electron beam is not illuminated on Ag but right next to Ag TEM specimen, we found that there is almost no oxidation. This suggests that ionization of residual $O_2$ molecules in the chamber does not play a major role for the oxidization, instead the $O_2$ molecules adsorbed on the Ag surface does. It is often believed that ionization of adsorbed $O_2$ is primarily introduced by secondary electrons instead of primary electrons, so the smallest structures fabricated by EBID have a typical width of 15–20 nm.[47] However, the low oxygen concentration in TEM chamber results in the localized oxidation within the irradiated area.[28] In principle, higher dose rate would create more active species and thus speed up the growth of $Ag_2O$ crystal if the oxidation is the only reaction happening under electron-beam irradiation. However, the irradiation also stimulates the desorption of O from the $Ag_2O$ surface via Auger electron excitation.[48–50] Irradiated by the incident electron-beam, $Ag^+$ can be further ionized and lose a core electron, followed by the decay of a valence electron from O to fill the resulting hole, which releases sufficient energy to excite another two electrons from $O^{2-}$: one to fill the core hole on the Ag cation and one ejected as an Auger electron, or both of them ejected. In such way, $O^{2-}$ turns into O or $O^+$, which will be repulsed by the Coulombic field from the $Ag^+$ ions and desorbed from the surface, resulting in the reduction of $Ag_2O$.[50] Since both the electron-beam ionization induced oxidation and O-depletion induced reduction are dose rate (or current density) dependent, whether $Ag_2O$ grain grows or decomposes, is determined by the competition between them. It is speculated that in current experimental set-up, (1) electron-beam irradiation is essential for the nucleation of $Ag_2O$; (2) both the oxidation rate and the reduction rate increase as the dose rate increases; (3) the oxidation rate has a higher start value while the reduction rate increases much faster.

In summary, the reversible oxidation and reduction processes in $Ag$-$Ag_2O$ are directly controlled by the dose rate of electron-beam. HRTEM observation indicates that oxygen is incorporated and extracted along the {111} planes in FCC Ag. It is speculated that the electron-beam induced ionization of $O_2$ molecules and the desorption of O from $Ag_2O$ play the major roles in the oxidation and reduction, correspondingly. In addition, the site-specific nano-scale metal-semiconductor heterostructures, as well as the $Ag_2O$ nano-arrays, can be realized by precisely configuring the electron-beam shape, size, position, and irradiation time. These findings do not only facilitate the basic understanding of oxidation/reduction kinetics in $Ag$-$Ag_2O$, but also open up a promising approach for precisely fabrication of nanostructures with metal or semiconductor properties in devices.

**Supporting Information.**

Supplementary Table S1, Figures S1 and S2 (PDF)

Movie S1 (AVI)

AUTHOR INFORMATION

**Corresponding authors**

*E-mail: (J. Wang) wang@whu.edu.cn.

*E-mail: (J. Wen) jwen@anl.gov.

**Author Contributions**

The manuscript was written through contributions of all authors. All authors have given approval to the final version of the manuscript. H. S. and H. Z. contributed equally to this work.

**Notes**

The authors declare no competing financial interest.

ACKNOWLEDGMENTS

Use of the Center for Nanoscale Materials, an Office of Science user facility, was supported by the U. S. Department of Energy, Office of Science, Office of Basic Energy Sciences, under Contract No. DE-AC02-06CH11357. This work was supported by the National Natural Science Foundation of China (51671148, 51271134, J1210061, 11674251, 51501132, and 51601132), the Hubei Provincial Natural Science Foundation of China (2016CFB446 and 2016CFB155), the Fundamental Research Funds for the Central Universities, the CERS-1-26 (CERS-China Equipment and Education Resources System), the China Postdoctoral Science Foundation (2014T70734), the Open Research Fund of Science and Technology on High Strength Structural Materials Laboratory (Central South University), and the Suzhou Science and Technology project (No. SYG201619).

ABBREVIATIONS

HRTEM, high-resolution transmission electron microscopy; ESD, electron stimulated desorption; FFT, fast Fourier transformation.